\documentclass[12pt,preprint]{aastex}

\newcommand{\msun}{\ensuremath{M_{\odot}}}	
\newcommand{\solarmass}[1]{#1\,\msun} 

\shorttitle{Very massive stars}
\shortauthors{Belkus et al.}

\begin{document}


\title{The evolution of very massive stars}


\author{H. Belkus\altaffilmark{1}, J. Van Bever \altaffilmark{2}, D. Vanbeveren\altaffilmark{1} \altaffilmark{3}}


\altaffiltext{1}{Astrophysical Institute, Vrije Universiteit Brussel, Pleinlaan 2, 1050 Brussels, Belgium. email: hbelkus@vub.ac.be , dvbevere@vub.ac.be}
\altaffiltext{2}{ Institute for Computational Astrophysics, Saint Mary's University, Halifax, NS,  B3H 3C3, Canada. email: vanbever@ap.smu.ca}
\altaffiltext{3}{ Mathematics department, Groep T, Vesaliusstraat 13, 3000 Leuven, Belgium.}


\begin{abstract}
Core collapse of dense massive star clusters is unavoidable and this leads to the formation of
massive  objects, with a mass up to 1000 $\msun$ and even larger. When these objects become stars, stellar
wind mass loss determines their evolution and final fate, and decides upon whether they form black holes
(with normal mass or with intermediate mass) or explode as a pair instability supernova. In the present
paper, we discuss the evolution of very massive stars and we present a convenient evolution recipe that
can be implemented in a gravitational N-body code to study the dynamics of dense massive clusters.
\end{abstract}


\keywords{stellar evolution - stellar wind - very massive stars}



\section{Introduction}  

The inner 100 pc of the Galactic center contains several young dense star clusters 
\citep{fig99a}, some of them with reliable mass estimates \citep{bor05}. Of particular
interest are the Arches cluster \citep{fig02}, the Quintuplet cluster  \citep{fig99b}, IRS
13E \citep{mai04} and IRS 16SW \citep{lu05}. Gravitational N-body simulations reveal that
soon after birth such clusters may experience core collapse where, depending on the initial cluster
radius, many or most of the massive stars participate in a "collision runaway" or "collision merger"
\citep{qui90,por99,fig02,gur04,gur05,frei06}. In a recent paper, \citet{por06}  estimated
the typical mass of these objects. They concluded that clusters in the inner 10 pc (respectively between
10 pc and 100 pc) of the Galactic center form collision mergers with an average mass $\approx$ 1000 $\msun$  
(respectively $\approx$ 500 $\msun$).  The dynamical evolution of a cluster where core collapse happens
obviously depends on whether or not the very massive merger becomes a very massive star and, when a very
massive star is formed, on the evolution of this very massive star. 

The present paper deals with the evolution of very massive stars, products of runaway merging. The
computation method is outlined in section 2 where we provide an easy evolution calculation recipe.
 It  is obvious that this evolution is critically affected by stellar wind mass loss. The formalism
that we use is discussed in section 3. The evolution of stars with a post-merger mass between 300 $\msun$
and 1000 $\msun$ is illustrated in section 4.

\section{Simulating the evolution of very massive stars} \label{section2}
In the subphotospheric layers of very massive stars, where the
opacity becomes larger than the electron scattering value, the radiation
force almost balances gravity, causing a core/extended halo stellar
structure. This hampers the convergence of stellar evolutionary
computations. However, since the mass of these layers is very small, they
hardly affect the overall internal structure and treating these layers using Thomson scattering opacity
only, still provides an accurate description of very massive stellar
evolution while at the same time avoiding any numerical difficulties. 

Our calculations of the evolution of very massive stars are based on the
results of \citet {nad05} (NR) who constructed interior
models for massive objects using the similarity theory of stellar
structure (treated as a boundary-value problem). Their models correspond
to chemically homogeneous stars, having Thomson scattering as the only
opacity source throughout. The obtained model sequences depend on one
parameter only; $\mu^2 M$ ($\mu$ being the mean molecular mass of the
gas and $M$ the total mass of the star). During most of their evolution very massive stars
produce a convective core that almost covers the entire star, meaning
that their evolution can be simulated accurately with a homogeneous
model. The fact that very massive stars are expected to lose a significant
amount of mass by stellar wind (section 3) strengthens the conclusion that very massive stars evolve
in a quasi-homogeneous way.  Furthermore, in case of steep dependencies of the energy generation rate on
temperature (as is the case for the CNO cycle and the $3\alpha$-reaction), most of the energy
production is localised near the very centre of the star. Therefore, to a
very good approximation, the luminosity is constant throughout the star
and the dimensionless luminosity equation is decoupled from the rest of
the stellar structure equations. This means that the model sequence of NR
can be used to describe the core hydrogen burning (CHB) as well as the core helium burning (CHeB) stage of very
massive stars.

To simulate the evolution of a star, we proceed as follows. NR provide
best fit relations (see their eqs. 30, and 34 in combination with 36) as a
function of $\mu^2 M$ for (a.o) the luminosity and convective core mass of
their computed sequence (which covers the range $0<\mu^2 M\leq
\solarmass{4000}$). For a given stellar luminosity (and
assuming that central nuclear burning is the only energy source in
the star), conservation of energy allows one to derive the
amount of nuclear fuel that is burned per unit of time.
Then, from knowledge of the size of the convective core, a differential
equation for the variation of the central abundance of the fuel is
obtained. For the CHB case, we have:

\begin{equation}
 M_{cc}(\mu, M)\,\frac{dX}{dt}= -\frac{L(\mu, M)}{\epsilon_H}
\end{equation}

\noindent where $\epsilon_H$ is the energy produced from burning one mass unit of
hydrogen. In this equation, both $M_{cc}$ and $L$ vary due to changes not
only of $\mu$, but also of $M$ through stellar wind mass loss:

\begin{equation}
 \frac{dM}{dt}= \dot{M}(\mu, M, L).
\end{equation}

Assuming the mass loss rate $\dot{M}$ can be derived from knowledge of
$\mu$, $M$ and/or $L$, the solution of this coupled set of two
differential equations provides the evolution of the quantities $X$,
$\mu$, $M$, $M_{cc}$, $L$ and $\dot{M}$ as a function of time, as well as
the duration of the Hydrogen burning stage (which ends, of course, when
$X=0$).

For the CHeB case, one needs to account for the fact that
C and O are produced in a non-constant ratio, which affects the energy
production per unit mass of burned helium. The equivalent of eq. 1
for Helium burning is (see also \citet {lan89a}): 

\begin{equation}
 M_{cc}(\mu,M)\left(\frac{B_Y}{A_Y}\frac{dY}{dt} +
\frac{B_C}{A_C}\frac{dC}{dt} + 
\frac{B_O}{A_O}\frac{dO}{dt}\right) = -L(\mu,M).
\end{equation}

In eq. 3, the $B$ symbols represent the binding energies of the
corresponding nuclei whereas the $A$ symbols are their atomic weights.
We impose the additional contraint $Y+C+O=1$ and thus $dY+dC+dO=0$, for
simplicity.

Langer (1989b) computed models for massive homogeneous CHeB
stars and found an abundance evolution of C and O relative to He that was
closely followed by all models ($\solarmass{15}\leq M
\leq\solarmass{100}$), independent of initial mass. This resulted in a fit
between C and Y (his eq. 1, see also his Fig. 1). According to Langer,
this fit is very accurate for $Y$ values larger than about 0.5, when
the O abundance is low and the $^{12}$C$(\alpha,\gamma)^{16}$O reaction is
relatively unimportant. At lower values of $Y$, this reaction and the
fact that some $^{16}$O is converted into $^{20}$Ne towards the end of
CHeB of very massive stars, produce an estimated uncertainty in the
fit of about 5-10\%. Using this $C(Y)$ relation and eliminating the $C$
and $O$ time derivatives from eq. 3, one finally obtains:

\begin{equation}
 M_{cc}(\mu,M)\left(
           \left( \frac{B_Y}{A_Y}-\frac{B_O}{A_O} \right) +
				       \left( \frac{B_C}{A_C}-\frac{B_O}{A_O} \right) C'(Y)
\right)\frac{dY}{dt} = -L(\mu,M).
\end{equation}

Here $C'(Y)$ denotes the derivative of the $C(Y)$ fit of Langer, with
respect to Y. As was the case for H burning, the combination of eq. 4 with
a mass loss rate formalism of the form of eq. 2 enables one to compute the
evolution of the star upto He depletion in the core.

In section 4 we will compare evolutionary results of massive stars which are calculated with
the similarity theory with results calculated with detailed stellar evolutionary codes in order to evaluate
our computational method.

\section{The stellar wind mass loss formalism of very massive stars} \label{section3}

Since direct observations of very massive stars in general, their stellar wind mass loss rates in particular
are lacking, we are forced to estimate the effect of stellar wind mass loss on very massive star evolution
either by extrapolating empirical formalisms holding for massive stars or we can use theoretical models when
they are available. \citet {kud02} studied line-drive winds of very massive stars and calculated mass loss
rates as function of metallicity Z of very massive O-type stars with a luminosity Log L/$L_{\odot}$ between 6.3
and 7.03 and for T$_{eff}$-values between 40000K and 60000K and he presented $\dot{M}$-interpolation formulae
for three different T$_{eff}$-values. Notice that the mass loss rates are smallest for the highest T$_{eff}$. 
This means that if, by using the $\dot{M}$-interpolation formula corresponding to the highest T$_{eff}$ during
the whole CHB phase, we predict that stellar wind mass loss of very massive stars is large, the real mass loss may
be even larger. 

Due to stellar wind mass loss during CHB, the post-CHB (CHeB) remnants of the very massive stars are hydrogen
deficient and may be considered as very massive Wolf-Rayet (WR) stars, e.g. the evolution of very massive
stars during CHeB has to be calculated accounting for WR-like mass loss rates.
Theoretical formalisms have been presented by \citet {Nug02}  but they contain stellar wind
parameters which can only be derived by linking a full hydrodynamical wind model to a stellar
evolutionary model (e.g. the temparature and radius in the wind at the sonic point, the wind velocity at infinity
etc). Therefore we prefer to rely on empirical formalisms. Based on indirect arguments involving population
synthesis of WR stars in the Solar neighbourhood (the WN/WC number ratio), on the masses of black holes in
binaries and on direct mass loss rate determinations of WR stars including the effects of clumping (before 1998 the
effects of clumping on empirical mass loss rates was investigated for only a few WR star), \citet {van98} (see also \citet {vanb03}) proposed the following relation

\begin{equation}
 Log(\dot{M}) = Log(L) - 10 + 0.5Log(Z/Z_{\odot}).
\end{equation}

\noindent where Z stands for the {\it initial} metallicity which is proportional to the Fe abundance of the WR star.

\citet {Nug00} used clumping-corrected mass loss loss rates of a large sample of Galactic WR stars
and proposed the following $\dot{M}$ formula as function of luminosity and helium abundance Y 

\begin{equation}
 Log(\dot{M}) = -11 + 1.29Log(L) + 1.7Log(Y) +0.5Log(1-X-Y).
\end{equation}

\noindent In the two formulae given above $\dot{M}$ is in $M_{\odot}$/yr and L in $L_{\odot}$.

\medskip

\noindent {\it {Remarks.}} \citet {kud02} calculated the mass loss rates for stars with a luminosity
$Log(L/L_{\odot}) $ up to 7. Initially (on the zero age main sequence) our 1000 $\msun $ star has $Log(L/L_{\odot} )
= 7.5$ and we extrapolated the 
$\dot{M}$-interpolation formulae. We obviously assured that the mass loss rates in the extrapolation zone are
smaller than or equal to the maximum mass loss rates for line driving as discussed by \citet {owo04}. 

The WR stars where both empirical WR mass loss rate formulae given above hold, have luminosities in the range
5.0 $\le$ LogL $\le$ 6.0. The theoretically predicted very massive WR stars (section 4) have luminosities up to
Log L = 7. The very massive WR mass loss rates that we use here are therefore extrapolated values implying
quite some uncertainty. In the next section we will discuss the consequences of this uncertainty. Similarly as for
the CHB mass loss rates it is obvious that also here we check that the extrapolated values remain smaller than or equal to
the maximum rates.       

\section{Results}

In order to illustrate to what extent our very massive star evolutionary scheme approaches detailed evolutionary
computations, Table 1 compares the 120 $\msun $ evolutionary result of \citet {sch92} with our
prediction where we obviously used the same mass loss prescription during CHB and during CHeB as in Schaller et
al. As can be noticed, the correspondence is very good. The basic assumption of our method is the
quasi-homogeneous evolution of very massive stars. The more massive a star, the larger is the convective
core and the larger is the stellar wind mass loss rate. This means that the more massive a star the closer its
evolution will be to the quasi-homogeneous one. Since our method gives very satisfactory results
already for the 120 $\msun $ star we are inclined to conclude that it will closely describe the evolution of very
massive stars. The latter is strengthened by the following. \citet {Mar03} calculated the evolution of a
1000 $\msun$ zero-metallicty star which is subject to a large stellar wind mass loss (they also use the Kudritzki
formalism). In Table 1 we also compare the results of Marigo et al. with ours whereas Figure 1 compares
the temporal behaviour of the mass of the convective core, the luminosity and the effective temperature. As can be noticed, the
correspondence is excellent.

We calculated the evolution of stars with a mass in the range 300-1000 $\msun $ for three metallicities, Z =
0.04, Z = 0.02 and Z = 0.001, using the core hydrogen burning $\dot{M}$ interpolation formulae corresponding to
the highest T$_{eff}$ (section 3). The collision of massive stars in a dense cluster happens typically 1 or 2 Myrs
after their formation. The central hydrogen abundance of the stars at the moment of collision may be
significantly smaller than the initial value which implies that even when the merger product is well mixed and
becomes homogeneous, the new X may be significantly smaller than the initial value of the cluster. For this
reason we performed evolutionary calculations of very massive stars  with X = 0.68, 0.6 and 0.5. The results are
summarized in Table 2, the initial mass - final mass relationship is depicted in Figure 2 (in this
figure we also plot the relation for stars with initial mass smaller than 100
\msun, taken from Van Bever and Vanbeveren, 2003), the temporal evolution of the stellar mass is shown in
Figure 3 and the evolutionary behaviour in a mass-luminosity diagram is given in Figure 4. They
illustrate the following conclusions:

\begin{itemize}

\item Very massive stars with Z $\ge$ 0.02 and with initial mass $\ge$ 300 $\msun $ lose most of their mass in the
form of stellar winds during CHB and CHeB. The same applies for very massive stars with Z = 0.001 and with initial
mass $\ge$ 500 \msun.

\item The final masses at the end of CHeB calculated with the WR mass loss rate formula 5 and 6 are very similar.
 
\item All the very massive stars with the same initial chemical composition and with an initial mass $\ge$ 300
$\msun $ end their life as stars with very similar final mass and their CHB and CHeB lifetimes are very similar.
This is the reason why in Table 2 we only give the details for the 300 $\msun $ and 1000 $\msun $ stars. 
  
\item A very massive star with a lower initial X has a shorter CHB lifetime, but a larger luminosity, thus a
higher stellar wind mass loss rate. This explains why the final masses for very massive stars with the same
initial metallicity Z hardly depend on the initial X.

\item Very massive OB-type stars with Z $\ge$ 0.02 and WR stars with Z $\ge$ 0.001 obey a very tight mass luminosity
relation, i.e.

\begin{equation}
 Log(L) = 1.07(Log(M))^2 - 4.62Log(M) + 11.8 \quad Z = 0.04
\end{equation}

\begin{equation}
 Log(L) = 1.12(Log(M))^2 - 4.98Log(M) + 12.4 \quad Z = 0.02
\end{equation}

for the OB stars and

\begin{equation}
	Log(L) = 1.23Log(M) + 4.19 \quad Z = 0.04
\end{equation}

\begin{equation}
	Log(L) = 1.14Log(M) + 4.37 \quad Z = 0.02
\end{equation}

\begin{equation}
	Log(L) = 0.88Log(M) + 4.98 \quad Z = 0.001
\end{equation}

for the WR stars. All relations have a determination coefficient R$^2$ $\ge$ 0.99.

\item When Z is larger than or equal to 0.02 our computations reveal that the very massive stars will end their
life as a $\le$ 40-50 $\msun $ black hole. Since the Galactic bulge has such a large Z, intermediate mass black
holes with a mass of a few 100 $\msun $ may be difficult to form there.  

\item For Z = 0.001 the final mass of very massive stars $\le$ 170 \msun. Therefore, intermediate mass black
holes (but with a mass of a few 100 \msun) may form in dense metallicity poor clusters.  

\item From the results of \citet {heg02} we conclude that when Z is between 0.001 and 0.02, one may
expect pair-instability supernova candidates among collision runaway mergers in clusters in the Galactic
center.
 
\end{itemize}

\noindent {\it {Remarks.}} The evolutionary computations discussed above let us conclude that it may
be difficult to form intermediate mass black holes by means of runaway merging in dense clusters in the Galactic
bulge (where Z $\ge$ 0.02). Of course the computations rely on the adopted mass loss rate formalisms. During
CHB we used the theoretically calculated $\dot{M}$-interpolation formula corresponding to the highest
T$_{eff}$-values which means that in reality the overall mass that is lost during CHB may be larger than
the values in Table 2. Note that \citet {kud02} calculated the mass loss rates of radiaton
driven stellar winds. Additional processes (like rotation for example) will increase the derived rates.
Furthermore, very massive stars may experience a luminous blue variable (LBV) phase somewhere near the end of
CHB, much like the massive stars do. The LBV phase is characterised by eruptive (explosive, \citet {smi06}) mass loss episodes (as observed in $\eta$ Car) and this may increase the total mass loss. The remarks
discussed above stengthen the main thesis of the present paper. 

A major uncertainty is obviously the (extrapolated) empirical WR mass loss rate formalism for the very massive
stars. To illustrate the importance of this uncertainty we computed the evolution of the very massive stars (Z
$\ge$ 0.02) but with WR mass loss rates which are a factor 2 and 4 smaller than predicted by the relations given
in section 3. The results are given in Table 2 as well. As expected, the final CHeB masses are larger and some of
them fall in the range where we expect pair-instability supernovae to happen, e.g. these stars do not form BHs at
all. Since a pair-instability supernova happens roughly when the final CHeB mass is larger than 65-75 \msun, we conclude from our computations and the argumentation above that it is very improbable that very massive stars in
the Galactic center produce black holes with a mass larger than 65-75 \msun.

\section{Summary}

In the present paper we studied the quasi-homogeneous evolution of very massive stars with mass up to 1000
$\msun$ which could be the result of core collapse of young dense clusters. When the theory of radiatively driven
stellar wind mass loss applies it follows that the evolution of very massive stars is dominated by these winds
 during CHB and during CHeB. At Solar metallicity and larger very massive stars end their live as a black hole
with a mass less then 75 \msun.  At Z = 0.001, the final mass of the very massive stars studied here may be a
factor 2 to three larger compared to those at Z = 0.02, e.g. in low metallicity regions the formation of
intermediate mass black holes with a mass of a few hundred $\msun $ is a possibility.  Furthermore it is very
plausible that between Z = 0.02 and Z = 0.001 at least some very massive stars will end their life with a
pair-instability supernova. During a pair-instability supernova very large amounts of metals may be ejected and
we like to suggest that the metal poor (Z $<$ 0.02) chemical evolution of galactic bulges may be affected by
cluster dynamics through the formation of very massive stars and the occurence of pair-instability supernova.

\medskip   

\noindent {\bf{Acknowledgement}}

\medskip
\noindent We like to thank an unknown referee for very valuable suggestions that improved the scientific
content of the paper.

\clearpage

\begin{figure}
\begin{center}
\plottwo{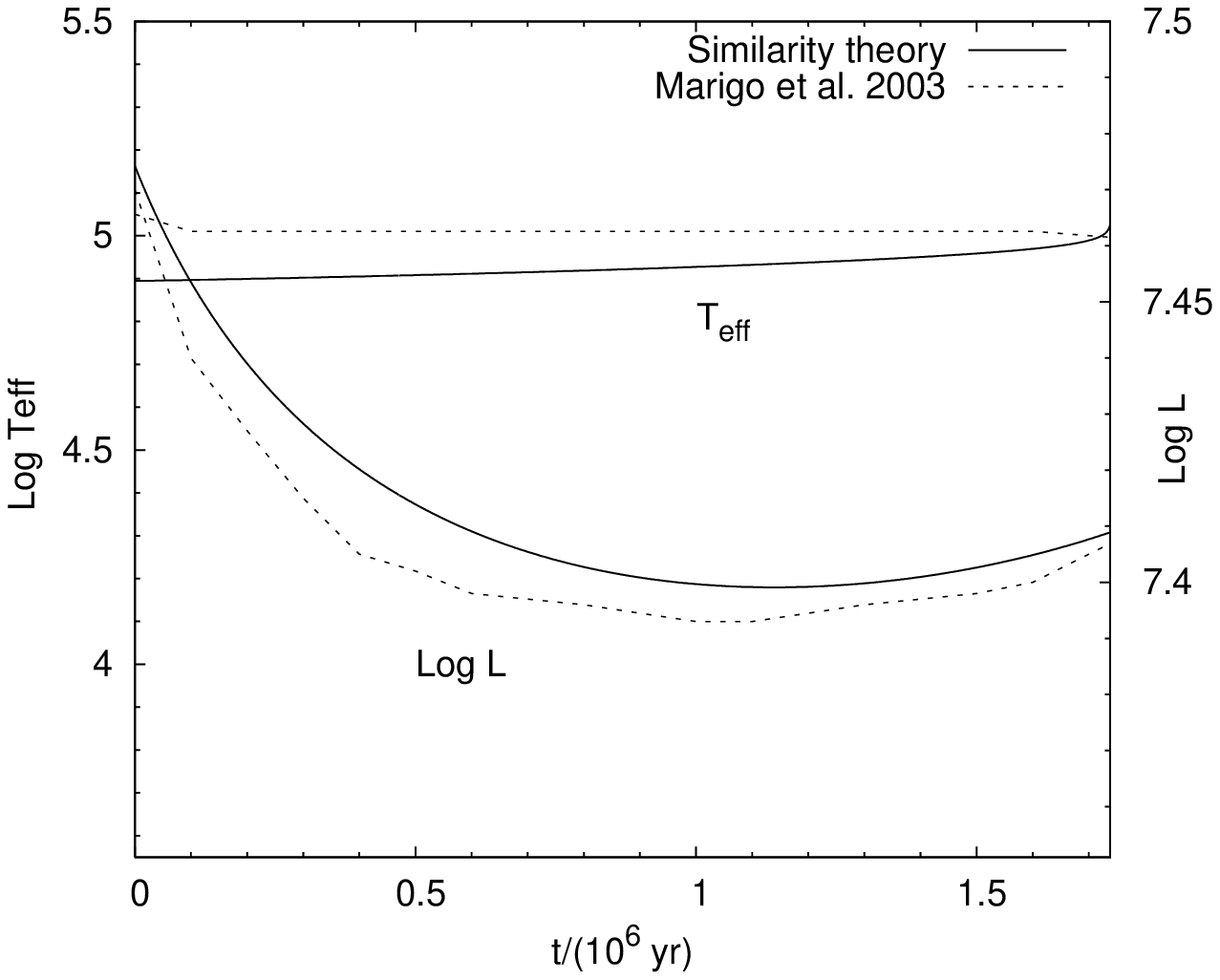}{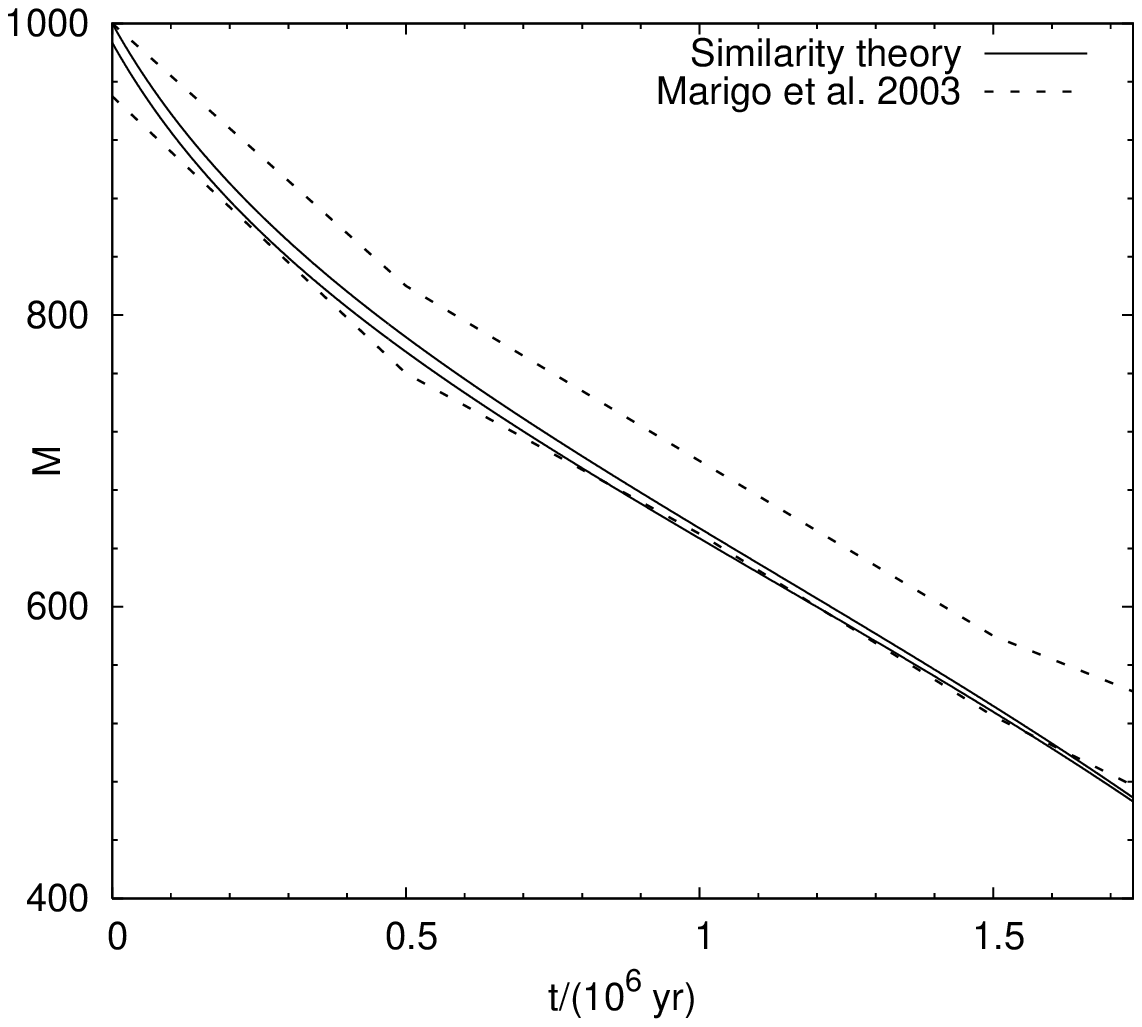}
\end{center}
\caption{Left figure: the temporal behaviour of the luminosity (in $L_{\odot}$) and of T$_{eff}$ during CHB of a zero
metallicity 1000 $\msun $ star: a comparison between the results of \citet {Mar03} (dashed lines) and our method
(full lines). Right figure: the same as the left figure but for the total mass (upper lines) and the mass of the convective core (lower lines).
(masses are in \msun). }
\end{figure}

\clearpage

\begin{figure}
\begin{center}
\epsscale{.60}
\plotone{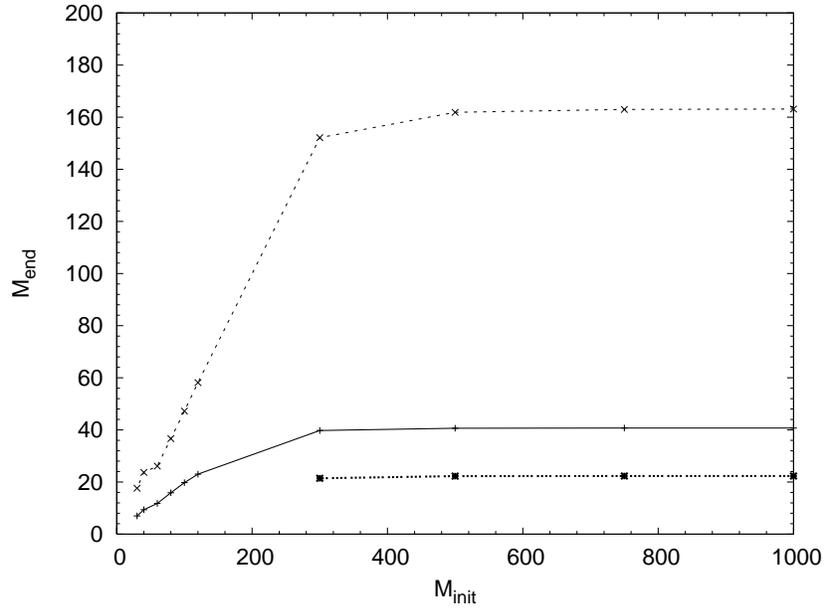}
\end{center}
\caption{The initial mass (M$_{init}$) - final mass (M$_{end}$) relation (all masses are in \msun) for Z = 0.04 (dotted),
Z = 0.02 (full line) and Z = 0.001 (dashed line).}
\end{figure}

\clearpage

\begin{figure}
\begin{center}
\includegraphics{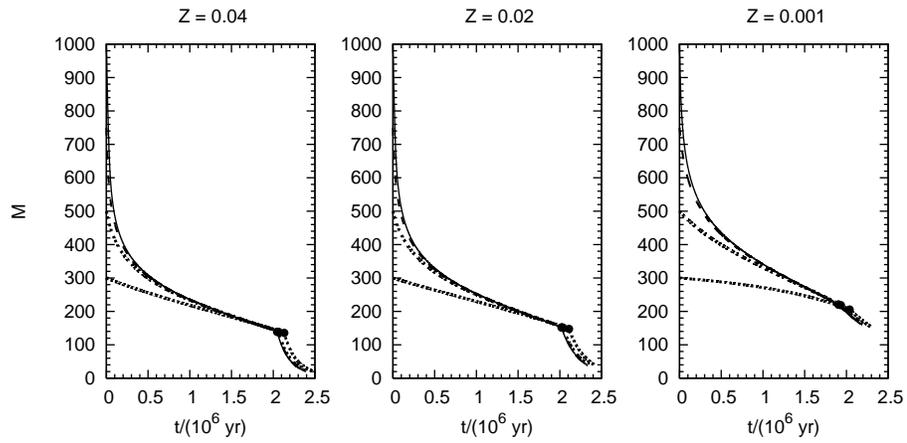}
\end{center}
\caption{The temporal evolution of the mass (in \msun) for the stars with initial mass = 1000 \msun, 750 \msun, 500 \msun
and 300 \msun. The dot-mark corresponds to the end of CHB = the beginning of the WR phase.}
\end{figure}

\clearpage

\begin{figure}
\begin{center}
\includegraphics{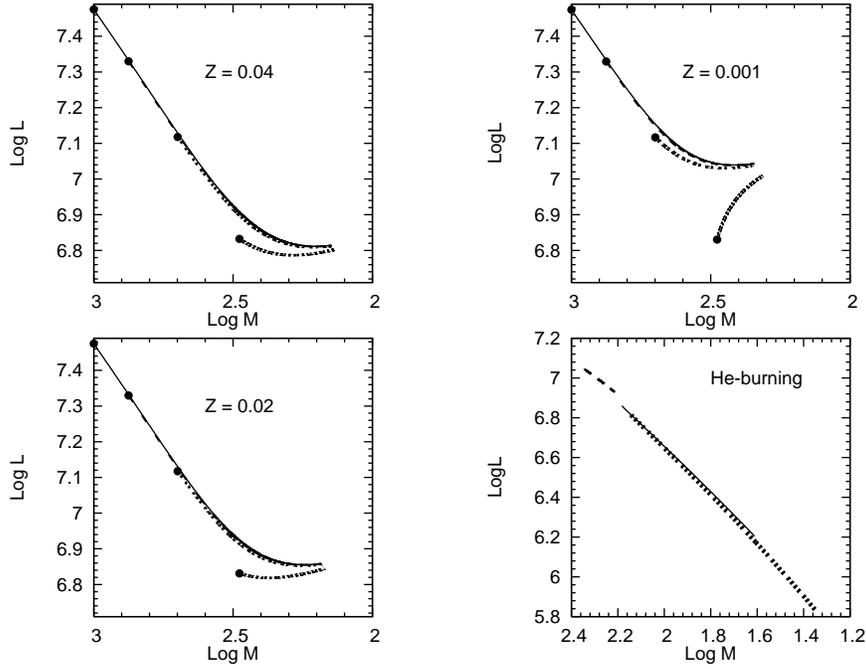}
\end{center}
\caption{The evolution in the mass - luminosity diagram (both in solar units) of the stars with initial mass = 1000 \msun,
750 \msun, 500 \msun and 300 \msun. The dots give the location of the initial zero age parameters of the four stars. The lower right figure shows the mass - luminosity evolution during the He-burning phase for the 3 metallicities: Z = 0.001 (dashed line), Z = 0.02 (full line) and Z = 0.04 (dotted line). }
\end{figure}

\clearpage

 \begin{table}
 \begin{center}
  \begin{tabular}{|c|c|l|l|l|l|c|}	\hline     
   \emph{$M_{init}$}& \emph{Z} & \emph{$M_{eCHB}$} & \emph{$T_{CHB}$}  & \emph{$M_{eWR}$}    &  \emph{$T_{CHeB}$}  &
\emph{reference} \\  \hline\hline
   
1000  & $2.10^{-6}$       & 466.76          & 2.050         & 415.18         & 2.31          &This paper  \\
            &                            & 482                 & 1.861        & 419               & 2.15          & \citet {Mar03}
\\
 120   &     0.02              & 69.64                & 2.753         & 7.91               & 4.65        &This paper  \\
           &                           & 70.28                & 2.611         & 7.77               & 4.40        & \citet {sch92}
\\        
                                              
 \hline
\end{tabular}
 \end{center}
 \caption{A comparison of massive and very massive star evolutionary computations performed with a detailed evolutionary
code and with the similarity method used in the present paper. All masses are in \msun, $T_{CHB}$ is in Myr, $T_{CHeB}$ is
in 10$^5$ yr.}
 \end{table}

\clearpage

\begin{table}
 \begin{center}
  \begin{tabular}{|c|c|c|c|c|l|l|}	\hline
   \emph{$M_{init}$}& \emph{Z} & \emph{X} & \emph{$M_{eCHB}$} & \emph{$T_{CHB}$}  & \emph{$M_{eCHeB}$}    & 
\emph{$T_{CHeB}$}  
\\  \hline\hline
   1000& 0.04 & 0.68       & 140.09          & 2.045        & 22.31  -  54.79  -  87.12           & 3.34  -  2.95  -  2.83  \\
          &          & 0.60       & 141.64          & 1.724         & 22.54                                                 & 3.33   \\
          &          & 0.50       & 144.35          & 1.353         & 22.95                                                 & 3.31   \\

          & 0.02 & 0.68       & 152.91          & 2.017        & 40.72  -  78.37  -  109.06  -  35.45       & 3.03  -  2.83  -  2.76  - 3.36\\
          &          & 0.60       & 154.56          & 1.701         & 41.19                                                 & 3.02   \\
          &          & 0.50       & 157.35          & 1.337         & 41.86                                                 & 3.01   \\

         & 0.001& 0.68       & 220.93          & 1.903        & 163.13                                                 & 2.62      \\
          &          & 0.60       & 222.47          & 1.610         & 164.41                                                 & 2.61   \\
          &          & 0.50       & 225.15          & 1.272         & 166.42                                                 & 2.61   \\      
                      
                             & &    &           &       &        &   \\
                             
  300 & 0.04 & 0.68       & 136.85          & 2.131        & 21.86  -  53.61  -  85.14           & 3.36  -  2.96  -  2.84 \\
          &          & 0.60       & 138.37          & 1.786         & 22.54                                                 & 3.33   \\
          &          & 0.50       & 140.95          & 1.394         & 22.95                                                 & 3.31   \\

          & 0.02 & 0.68       & 148.76          & 2.109        & 39.77  -  76.30  -  106.09  -  34.86         & 3.04  -  2.84  -  2.77  - 3.38\\
          &          & 0.60       & 150.20         &  1.769        & 40.05                                                          & 3.04  \\
          &          & 0.50       &  152.74        &   1.382       & 40.69                                                          &  3.03 \\

         & 0.001& 0.68       & 205.96          & 2.038        & 152.15                                                 & 2.64      \\
          &          & 0.60       & 207.30          & 1.710         & 153.09                                                 & 2.64   \\
          &          & 0.50       & 209.63          & 1.337         & 154.96                                                 & 2.63   \\

     \hline
  \end{tabular}
 \end{center}
 \caption{Evolutionary properties of very massive stars as function of initial chemical composition (Z, X). We list the
mass at the end of CHB ($M_{eCHB}$), the CHB timescale ($T_{CHB}$), the mass at the end of CHeB ($M_{eCHeB}$) and the CHeB
timescale ($T_{CHeB}$). The CHeB numbers are always calculated using the WR mass loss rate formula 5.  For Z = 0.04, Z =
0.02 and X = 0.68 we also list the two CHeB parameters using WR mass loss rate formula 5 divided by 2 and
using formula 5 divided by 4. For Z = 0.02 and X = 0.68 the fourth number corresponds to the case where the WR mass loss
rate is computed with formula 6. All masses are in \msun, $T_{CHB}$ is in Myr, $T_{CHeB}$ is in 10$^5$ yr.}
 \end{table}
 

\begin{thebibliography}{}
\bibitem [Borissova et al.(2005)]{bor05}Borissova, J., Ivanov, V.D., Minniti, D., Geisler, D., Stephens, A.W.: 2005, \aap 435, 95.
\bibitem [Figer  \& Kim(2002)] {fig02}Figer, D.F., Kim, S.S.: 2002, in ASP Conf. Ser. 263: Stellar Collisions, Mergers and their Consequences, pp. 287.
\bibitem [Figer et al.(1999b)] {fig99b} Figer, D.F., McLea, I.S., Morris, M.: 1999b, \apj 514, 202.
\bibitem [Figer et al.(2002)] {fig02} Figer, D.F., Najarro, F., Gilmore, D., Morris, M., Kim, S.S., Serabyn, E., McLean, I.S., Gilbert, A.M., Graham, J.R., Larkin, J.E., Levenson, N.A., Teplitz, H.I.: 2002, \apj 581, 258.
\bibitem [Figer et al.(1999a)] {fig99a}Figer, D.L., Kim, S.S., Morris, M., Serabyn, E., Rich, R.M., McLea, I.S.: 1999a, \apj 525, 750.
\bibitem [Freitag et al. (2006)] {frei06}Freitag, M., G\"urkan, M.A., Rasio, F.A.: 2006, MNRAS 368, 141.
\bibitem [G\"urkan et al. (2004)] {gur04} G\"urkan, M.A., Freitag, M., Rasio, F.A.: 2004, \apj 604, 632.
\bibitem [G\"urkan \& Rasio (2005)] {gur05} G\"urkan, M.A., Rasio, F.A.: 2005, \apj 628, 236.
\bibitem [Heger \& Woosley (2002)] {heg02}Heger, A., Woosley, S.E.: 2002, \apj 567, 532.
\bibitem [Ishii  et al. (1999)] {ish99} Ishii, M., Ueno, M., Kato M.: 1999, PASP 51, 417.
\bibitem [Kudritzki (2002)] {kud02}Kudritzki, R.P.: 2002, \apj 577, 389.
\bibitem [Langer  (1989a)] {lan89a}Langer, N., A\&A, 1989a, 220, 135.
\bibitem [Langer  (1989b)] {lan89b}Langer N., A\&A, 1989b, 210, 93.
\bibitem [Lu et al. (2005)] {lu05} Lu, J.R., Ghez, A.M., Hornstein, S.D., Morris, M., Becklin, E.E., 2005, \apj 625, 51.
\bibitem [Marigo et al. (2003)] {Mar03}Marigo,ÊP., Chiosi,ÊC., Kudritzki,ÊR.-P.Ê2003, A\&A 399, 617.
\bibitem [Maillard et al. (2004)]{mai04}Maillard, J.P., Paumard, T., Stolovy, S,R., Rigaut, F.:,2004, A\&A 423, 155.
\bibitem [Nadyozhin  \& Razinkova (2005)] {nad05}Nadyozhin, D.K., Razinkova, T.L., 2005, Astronomy Letters, Vol. 31, No. 10, pp. 695.
\bibitem [Nugis \&  Lamers (2000)] {Nug00} Nugis, T., Lamers, H.J.G.L.M.: 2000, A\&A 360, 227.
\bibitem [Nugis \&  Lamers (2002)] {Nug02}  Nugis, T., Lamers, H.J.G.L.M.: 2002, A\&A 389, 162.
\bibitem [Owocki et al. (2004)] {owo04} Owocki, S.P., Gayley, K.G., Shaviv, N.J., 2004, \ apj, 616, 525.
\bibitem [Portegies Zwart et al. (2006)] {por06}Portegies Zwart, S.F., Baumgardt, H., McMillan, S.L.W., Makino, J., Hut, P., Ebisuzaki, T.: 2006, \apj (in press) (astro-ph/0511397).
\bibitem [Portegies Zwart et al. (1999)] {por99}Portegies Zwart, S.F., Mkino, J., McMillan, S.L.W., Hut, P.: 1999, A\&A 348, 117.
\bibitem [Quinlan  \& Shapiro (1990)] {qui90} Quinlan, G.D., Shapiro, S.L.: 1990, \apj 356, 483.
\bibitem [Schaller et al. (1992)] {sch92}Schaller, G., Schaerer, D., Meynet, G., Maeder, A.: 1992, A\&AS 96, 269.
\bibitem [Smith \&  Owocki (2006)] {smi06}Smith, N., Owocki, S.P.: 2006, \apj 645, L45. 
\bibitem [Van Bever \& Vanbeveren (2003)] {vanb03}Van Bever, J., Vanbeveren, D.: 2003, A\&A 400, 63.
\bibitem [Vanbeveren et al. (1998)] {van98}Vanbeveren, D., De Loore, C., Van Rensbergen, W.: 1998, The A\&A Rev. 9, 63. 

\end{thebibliography}
\end{document}